\documentstyle[11pt]{article}

\columnwidth\textwidth
\title{Non-abelian gauge theories as a consequence\\
of perturbative quantum gauge invariance  \thanks{Work
supported by Swiss National Science Foundation}}
\author{A. Aste and G. Scharf\\
Institut f\"ur Theoretische Physik der Universit\"at Z\"urich\\
Winterthurerstrasse 190, CH-8057 Z\"urich, Switzerland}
\date{}
\begin{document}
\maketitle

\begin{abstract}
We show for the case of interacting massless vector bosons,
how the structure of Yang-Mills theories emerges
automatically from a more fundamental concept,
namely perturbative quantum gauge invariance.
It turns out that the coupling in a non-abelian gauge theory
is necessarily of Yang-Mills type
plus divergence- and coboundary couplings. The extension of the method
to massive gauge theories is briefly discussed.
\\

{\bf PACS.} 11.10 - Field theory, 11.20 - S-matrix theory, 12.35 - General
properties of quantum chromodynamics.
\end{abstract}

\def\d{\partial}\def\=d{\,{\buildrel\rm def\over =}\,}
\def\sqr#1#2{{\vcenter{\vbox{\hrule height.#2pt\hbox{\vrule width.#2pt 
height#1pt \kern#1pt \vrule width.#2pt}\hrule height.#2pt}}}}
\def\w{\mathchoice\sqr45\sqr45\sqr{2.1}3\sqr{1.5}3\,} 

\newpage

\section{Introduction}
It is a well-known fact that gauge theories have many
pleasing features. Apart from the beauty of the mathematical
structure of the classical theories, it is the renormalizability
of their quantum versions which makes them so important for
particle physicists. What is less known is that some basic properties
of non-abelian quantum gauge theories can be understood as the
consequence of a suitable quantum version of gauge
invariance. It is the aim of this paper to demonstrate this fact
for the case of interacting massless vector bosons. The result will be
the usual Yang-Mills plus ghost couplings plus some divengence- and
coboundary couplings. The nice feature of this approach is that the
Lie algebra structure of the Yang-Mills theory is not put in, but comes 
out: the antisymmetry of the structure constants follows from first 
order gauge invariance and the Jacobi identity from second order.
This fact has been noticed by R. Stora in a more special
setting \cite{1}. The present method has already been used by us to analyse
the abelian Higgs model \cite{2} and the electroweak theory \cite{3,4}.

The paper is organized as follows: In section 2 we present
our general framework and define perturbative
quantum gauge invariance on the Fock space of free asymptotic massless
vector bosons and scalar fermionic ghosts \cite{5,6}. In section 3
we investigate how gauge invariance restricts all renormalizable couplings
of the gauge fields at first and second order of perturbation
theory. In section 4 we give a short outlook how the presented
method can be generalized to the case of massive gauge bosons.

\section{Theoretical Framework}
\subsection{Construction of the S-Matrix}
\vskip 0.5 cm
In this paper, we consider the perturbative
S-matrix as a sum of smeared operator-valued distributions of the following
form \cite{7,8}:
\begin{equation}
S(g)={\bf 1}+\sum \limits_{n=1}^{\infty} \frac{1}{n !} \int dx_1 ... dx_n
T_n(x_1,...x_n) g(x_1) \cdot ... g(x_n), \label{stoer}
\end{equation}
where $g\in\cal{S}$, the Schwartz space of functions of rapid
decrease.  
The test function $g$ plays the role of 'adiabatic switching'
and provides a cutoff in the long-range part of the interaction, without
destroying any symmetry. It can be considered as a natural infrared regulator.
We work with free asymptotic fields throughout, so that all expressions
are well-defined; interacting fields nowhere appear.

The $T_n(x_1,...x_n)$ are well-defined time-ordered products of the first
order coupling $T_1(x)$, they are expressed in terms of Wick monomials 
of free fields. The construction of the $T_n$ requires some care:
If the arguments $(x_1,...x_n)$ are all time-ordered, i.e.
if we have
\begin{equation}
x_1^0>x_2^0>...>x_n^0 \quad ,
\end{equation}
then $T_n$ is simply given by
\begin{equation}
T_n(x_1,...x_n) = T_1(x_1)T_1(x_2) ...\cdot T_1(x_n)
\end{equation} 
According to the definition (\ref{stoer})
$T_n(x_1,...x_n)$ is symmetric in $x_1,...x_n$. Using this fact
allows us in principle to obtain the operator-valued distribution
$T_n$ inductively
everywhere except for the complete diagonal $\Delta_n=\{
x_1=...=x_n\}$ [9].If $T_n$ were a C-number distribution, 
we could make it a well-defined
distribution for all $x_1,...x_n$ by extending the distribution
from $R^{4n}/\Delta_n$ to $R^4$.
The problem can be reduced to a C-number problem by the Wick
expansion of the operator-valued distributions.
The extension $T(x_1,...x_n)$ is, of course, not unique: it is ambiguous
up to distributions with local support $\Delta_n$. This ambiguity
can be further reduced by the help of symmetries (in particular gauge 
invariance) and power counting theory.

The concrete inductive construction of the $T_n$
has also to be performed carefully. Since the behaviour
of the distributions in $p$-space is much better than the very singular
one in $x$-space, it is advantageous to use the original Epstein-Glaser
method of splitting causal distributions \cite{8,10}, instead of the above 
extension method, because the former can be translated to $p$-space. 
The well-known ultraviolet divergences are in fact a consequence of 'careless'
splitting of operator valued distributions.

\subsection{Definition of Perturbative Quantum Gauge Invariance}
We introduce the concept of gauge invariance for 
the simple case of quantum electrodynamics first, where the coupling of
the electron to the photon is given at first order by
\begin{equation}
T_1(x) = ie:\bar{\Psi}(x) \gamma^\mu \Psi(x): A_\mu (x) \quad .
\end{equation}
Let
\begin {equation}
Q \stackrel{def}{=} \int d^3 x (\partial_\mu A^\mu(x) \stackrel
{\leftrightarrow}{\partial_0} u(x)) \label{Q}
\end{equation}
be the generator of (free) gauge transformations, called gauge charge for
brevity. This $Q$ has first been introduced in a paper by T. Kugo and 
I. Ojima \cite{11}.
$A_\mu$ is the gauge potential in the Feynman gauge,
and we choose $u$ as a scalar fermionic field, in order to have $Q^2=0$ 
(see section 2.3). The free fields satisfy the well-known
commutation relations
\begin{equation}
[A_\mu^{(\pm)}(x),A_\nu^{(\mp)}(y)]=ig^{\mu \nu} D^{(\mp)}(x-y),
\end{equation}
\begin{equation}
\{u^{(\pm)}(x),\tilde{u}^{(\mp)}(y)\}=-iD^{(\mp)}(x-y),
\end{equation}
and all other commutators vanish. $D^{\mp}$ are the well-known 
positive and negative frequency parts of the Pauli-Jordan distribution.
All the fields fulfil the
Klein-Gordon equation with zero mass, and as already mentioned, we are
working in Feynman gauge, but we are not forced to do so.
The following discussion would go through with some technical changes
in other covariant $\xi$-gauges as well [12].
In order to see how the
infinitesimal gauge tranformation acts on the free fields, we calculate
the commutators \cite{5}
\begin{equation}
[Q,A_\mu]=i\partial_\mu u \quad ,
\quad \{Q,u\}=0 \quad , \quad \{Q,\tilde{u}\}=-i\partial_\nu A^\nu \quad ,
\quad [Q,\Psi]=[Q,{\bar{\Psi}}]=0 \quad .
\end{equation}
Then we have
\begin{displaymath}
[Q,T_1(x)] = -e:\bar{\Psi} \gamma^\mu \Psi: \partial_\mu u
\end{displaymath}
\begin{equation}
=i \partial_\mu (ie:\bar{\Psi} \gamma^\mu \Psi:u) = i \partial_\mu T^\mu_{1/1}
(x). \label{gi}
\end{equation}
Note that the free electron field is {\em{not}} affected by the
gauge transformation.
We will call $T^\mu_{1/1}$ the 'Q-vertex' in the sequel.
The generalization of (\ref{gi}) to n-th order is
\begin{equation}
[Q,T_n(x_1,...x_n)] = i \sum_{l=1}^{n} \partial_\mu^{x_l} T_{n/l}^{\mu}
(x_1,...x_n) = (\hbox{\rm sum of divergences}) \quad , \label{dive}
\end{equation}
where $T^\mu_{n/l}$ is a mathematically rigorous version of the time-ordered
product
\begin{equation}
T^\mu_{n/l}(x_1,...,x_n)
\, "=" \, T(T_1(x_1)...T^\mu_{1/1} (x_l)...T_1(x_n)) \quad ,
\end{equation}
constructed by means of the methods described in section 2.1.
We define (\ref{dive}) to be the condition of gauge invariance [5].
For a fixed $x_l$ we consider from $T_n$ all terms with the external field
operator $A_\mu(x_l)$
\begin{equation}
T_n(x_1,...x_n) = :t^\mu_l(x_1,...x_n) A_\mu(x_l):+...
\end{equation}
(the dots represent terms without $A_\mu(x_l)$).
Then gauge invariance (\ref{dive}) requires
\begin{equation}
\partial_\mu^l [t^\mu_l(x_1,...x_n)u(x_l)]=t^\mu_l(x_1,...x_n) \partial_\mu
u(x_l)
\end{equation}
or
\begin{equation}
\partial_\mu^l t^\mu_l(x_1,...x_n) = 0 \quad , {\label{eich}}
\end{equation}
i.e. we obtain the Ward-Takahashi identities \cite{13} for QED.

As a further step, we consider the Yang-Mills theory without fermions now.
The first order coupling shall be given by
\begin{equation}
T_1(x)=igf_{abc} \{\frac{1}{2}: A_{\mu a}(x) A_{\nu b}(x) F^{\nu \mu}_c (x):
-:A_{\mu a}(x) u_b (x) \partial^{\mu} {\tilde{u}}_c(x): \} \quad , \label{QCD}
\end{equation}
where $F_a^{\nu \mu} = \partial^\nu A^\nu_a - \partial^\mu A^\nu_a$ is
the free field strength tensor, $u_a, \tilde{u}_a$ are the
(fermionic) ghost fields, and the $f_{abc}$ are the usual
$SU(N)$ structure constants.
The asymptotic free fields satisfy the commutation relations
\begin{equation}
[A_{\mu a}^{(\pm)}(x),A_{\nu b}^{(\mp)}(y)]=i \delta_{ab}
g_{\mu \nu} D^{(\mp)}(x-y)
\end{equation}
and
\begin{equation}
\{u_a^{(\pm)}(x),\tilde{u}_b^{(\mp)}(y)\}=-i \delta_{ab} D^{(\mp)}(x-y),
\end{equation}
and all other \{anti-\}commutators vanish. 
The introduction of ghost couplings is necessary here
to preserve perturbative quantum gauge invariance at first order.

Defining the gauge charge as in (\ref{Q}) by
\begin{equation}
Q:=\int d^3x \partial_\mu A^\mu_a (x)  \stackrel
{\leftrightarrow}{\partial_0} u_a(x) \quad ,
\end{equation}
where summation over repeated indices is understood,
we are led to the following (anti-) commutators with the fields:
\begin{equation}
[Q,A_a^\mu]=i\partial^\mu u_a \quad ,
\quad [Q,F_a^{\mu \nu}]=0 \quad , \quad
\{Q,u_a\}=0 \quad , \quad \{Q,\tilde{u}_a\}= -i\partial_\mu A^\mu_a \quad .
\label{comm}
\end{equation}
Obviously, these gauge
variations have a 'simpler' structure than the well-known
BRS transformation, because the non-abelian parts are missing here.
Calculating $[Q,T_1(x)]$ gives the Q-vertex
\begin{equation}
T^\mu_{1/1} = ig f_{abc}\bigl\{ 
 :u_a A_{\nu b} F^{\mu \nu}_c:
+ \frac{1}{2} : u_a u_b \partial^\mu \tilde{u}_c : \}
\end{equation}
and the first order coupling is gauge invariant through the presence of
a ghost coupling term in $T_1$.
The first order coupling given in (\ref{QCD}) is in fact not the most
general one, and it is also possible to construct a gauge invariant
first order coupling with bosonic ghosts \cite{5}; but gauge invariance
then breaks down at second order of perturbation theory. Gauge
invariance at order $n$ is again defined by (11).

Note that the usual 4-gluon term $\sim g^2$
is missing in $T_1$. This term appears
as a necessary local normalization term at second order, as we shall
see (\ref{viergl}).
This stresses the fact that our notion of perturbative gauge invariance
is strongly related to the formal expansion of the theory in powers of
the coupling constant g, in contrast to the conventional treatment
which discusses quantum effects of a local gauge group via formal
expansions in powers of $\hbar$. In our approach, no local gauge group
is assumed from the start. Local gauge groups will rather appear as a
consequence of perturbative gauge invariance.

\subsection{Important properties of the gauge charge Q}
Using the Leibnitz rule for graded algebras gives
\begin{displaymath}
Q^2=\frac{1}{2} \{Q,Q\} =\frac{1}{2} \int \limits_{x_0=const.}
d^3 x \partial_\nu A^\nu_a(x) \{  \stackrel
{\leftrightarrow}{\partial_{x_0}} u_a(x) , Q \}  \label{nr22}
\end{displaymath}
\begin{equation}
-\frac{1}{2}  \int \limits_{x_0=const.}
d^3 x [\partial_\nu A^\nu_a(x) , Q] \stackrel{\leftrightarrow}{\partial_{x_0}}
u_a(x) = 0 \quad ,
\end{equation}
i.e. Q is nilpotent. This basic property of Q and the Krein structure on the
Fock-Hilbert space \cite{14,15} allows to prove unitarity of the
$S$-matrix on the physical Hilbert space $H_{phys}$, which is a subspace of the
Fock-Hilbert space $F$ containing also the unphysical ghosts and
unphysical degrees of freedom of the vector field \cite{6}.

The physical Fock space can be expressed by the kernel and the
range of Q \cite{6,14}, namely
\begin{equation}
H_{phys}=ker \, Q / ran \, Q=ker\{Q,Q^+\} \quad . \label{physspace}
\end{equation}
This can be most easily seen by realizing the various
field operators on a {\em positive definite} Fock-Hilbert space $F$
as follows:
\begin{equation}
A_0(x)=(2 \pi)^{-3/2} \int \frac{d^3 p}{\sqrt{2 p^0}} [a_0(\vec{p})
e^{-ipx} - a_0^+(\vec{p}) e^{ipx}] \quad ,
\end{equation}
\begin{equation}
A_j(x)=(2 \pi)^{-3/2} \int \frac{d^3 p}{\sqrt{2 p^0}} [a_j(\vec{p})
e^{-ipx} + a_j^+(\vec{p}) e^{ipx}] \quad  \mbox{for} \quad j=1,2,3,
\end{equation}
\begin{equation}
u(x)=(2 \pi)^{-3/2} \int  \frac{d^3 p}{\sqrt{2 p^0}}
\Bigl(c_2(\vec p)e^{-ipx}+c_1^+(\vec p)e^{ipx}\Bigl)
\quad ,
\end{equation}
\begin{equation}
\tilde u(x)=(2 \pi)^{-3/2} \int  \frac{d^3 p}{\sqrt{2 p^0}}
\Bigl(-c_1(\vec p)e^{-ipx}+c_2^+
(\vec p)e^{ipx}\Bigl) \quad ,
\end{equation}
and
\begin{equation}
[a_\nu(\vec{p}),a_\mu^+(\vec{p})]=\delta_{\mu \nu} \delta(\vec{p}-\vec{q})
\quad , \quad \mu=0,1,2,3,
\end{equation}
\begin{equation}
\{c_i(\vec p),c_j^+(\vec q)\}=\delta_{ij}\delta(\vec p-\vec q),
\quad i,j=1,2,
\end{equation}
Above, $^+$ denotes the adjoint with respect to the positive scalar product
so that the operators can be represented in the usual way in the Fock space.
The Krein structure is then defined by introducing the conjugation $K$
\begin{equation}
a_0(\vec{p})^K = -a_0(\vec{p})^+ \quad ,
\quad a_j(\vec{p})^K=a_j(\vec{p})^+ \quad ,
\end{equation}
so that $A_\mu^K=A_\mu$, and on the ghost sector
\begin{equation}
c_2(\vec p)^K=c_1(\vec p)^+,\quad c_1(\vec p)^K=c_2(\vec p)^+ \quad ,
\end{equation}
so that $u^K=u$ is $K$-selfadjoint and $\tilde u^K=-\tilde u$. Then
$Q$ is densely defined and becomes $K$-symmetric $Q
\subset Q^K$.

The anticommutator in (\ref{physspace}) is essentially the number
operator for unphysical particles
\begin{equation}
\{Q^+,Q\} = 2 \int d^3 p \, \vec{p}^2 \bigl[ b_1^+(\vec{p}) b_1(\vec{p}) +
b_2^+(\vec{p})b_2(\vec{p})+c_1^+(\vec{p})c_1(\vec{p})+c_2^+(\vec{p})
c_2(\vec{p}) \bigr] \quad ,
\end{equation}
with
\begin{equation}
b_{1,2}=(a_{\|} \pm a_0)/\sqrt{2} \quad , \quad
a_{\|}=p_ja_j/|\vec{p}| \quad ,
\end{equation}
which implies (\ref{physspace}).

The nilpotency of $Q$ allows for standard homological notions \cite{6}:
Consider the field algebra $\cal{F}$ consisting of the polynomials in the
(smeared) gauge and ghost fields and their Wick powers. 
Defining a gauge variation for a Wick monomial $F$ according to
\begin{equation}
d_Q F \stackrel{def}{=} QF - (-1)^{n_F} FQ \quad ,
\end{equation}
where $n_F$ is the number of ghost fields in F,
Q becomes a differential operator in the sense of homological algebra,
and we have
\begin{equation}
d_Q^2=0 \Longleftrightarrow \{Q,[Q,F_b]\} = [Q,\{Q,F_f\}]=0 \quad ,
\end{equation}
where $F_b$ is a bosonic and $F_f$ a fermionic operator
and $d_Q(FG)=(d_QF)G+(-1)^{n_F}Fd_QG$.
For example, we get
\begin{equation}
d_Q :A_{\mu a} u_b \partial^\mu \tilde{u}_c: = :[Q,A_{\mu a}]u_b
\partial^\mu \tilde{u}_c: + :A_{\mu a}[Q,u_b] \partial^\mu \tilde{u}_c:
-:A_{\mu a}u_b \{Q, \partial^\mu \tilde{u}_c\}: \quad .
\end{equation}
If $F=d_QG$, then $F$ is called a coboundary.

\section{Implications of Perturbative Quantum Gauge Invariance}
\subsection{First Order Gauge Invariance}
A theory with massless fields suffers usually from serious infrared
divergences which may have deep relevance to the confinement
mechanism. This, however, is of no importance in our context, since
perturbative gauge invariance concerns the $T_n$ only. 

We describe the interaction of the massless 'gluons' by the most general
renormalizable ansatz with zero ghost number (coupling with non-zero
ghost number would affect the theory only for unphysical processes)
\begin{eqnarray}
\tilde{T}_1(x) & = ig \Bigl\{ 
& \tilde{f}^1_{abc} : A_{\mu a}(x) A_{\nu b}(x)
\partial^\nu A^\mu_c (x): +
\nonumber \\
& & \tilde{f}^2_{abc} : A_{\mu a} u_b \partial^\mu \tilde{u}_c : +
\nonumber \\
& & \tilde{f}^3_{abc} : A_{\mu a} \partial^\mu u_b \tilde{u}_c : +
\nonumber \\
& & \tilde{f}^4_{abc} : A_{\mu a} A^\mu_b \partial_\nu A ^\nu_c: +
\nonumber \\
& & \tilde{f}^5_{abc} : \partial_\nu A^\nu_a u_b \tilde{u}_c : \Bigr\}
\quad , \quad  \tilde{f}^4_{abc} = \tilde{f}^4_{bac} \quad ,\nonumber
\end{eqnarray}
where the $f$'s are arbitrary real constants. The
first order coupling term is then antisymmetric with respect
to the conjugation $K$: $\tilde{T}_1^K = -\tilde{T}_1$. 

Adding divergence terms to $\tilde{T}_1$ will not change the
physics of the theory (see sect.3.3), so that we always calculate modulo
divergences in the following. We can
modify $\tilde{T}_1$ by adding divergences $-\frac{ig}{4}
(\tilde{f}^1_{abc}+\tilde{f}^1_{cba}) \partial_\nu : A_{\mu a} A_{\nu b}
A^\mu_c :$ and $-ig \tilde{f}^3_{abc} \partial_\mu : A^\mu_a u_b \tilde{u}_c:$
and arrive at the equivalent first order coupling
\begin{eqnarray}
T_1 & = ig \bigl\{ 
& f^1_{abc} : A_{\mu a} A_{\nu b} \partial^\nu A^\mu_c: +
\nonumber \\
& & f^2_{abc} : A_{\mu a} u_b \partial^\mu \tilde{u}_c : +
\nonumber \\
& & f^4_{abc} : A_{\mu a} A^\mu_b \partial_\nu A^\nu_c : +
\nonumber \\
& & f^5_{abc} : \partial_\nu A^\nu_a u_b \tilde{u}_c : \bigr\} \quad ,
\label{nr35}
\end{eqnarray}
where 
\begin{equation}
f^1_{abc}=-f^1_{cba} \quad f^4_{abc}=f^4_{bac} \quad . \label{asy}
\end{equation}
The gauge variation of $T_1$ is
\begin{eqnarray}
d_Q T_1 & = -g f^1_{abc} & \Bigl\{
: \partial_\mu u_a A_{\nu b} \partial^\nu
A^\mu_c +
\nonumber \\
& & A_{\mu a} \partial_\nu u_b \partial^\nu A^\mu_c +
\nonumber \\
& & A_{\mu a} A_{\nu b} \partial^\nu \partial^\mu u_c : \Bigr\}+
\nonumber \\
& -g f^2_{abc} & \Bigl\{: \partial_\mu u_a u_b \partial^\mu \tilde{u}_c +
\nonumber \\
& & A_{\mu a} u_b \partial^\mu \partial_\nu A^\nu_c: \Bigr\} +
\nonumber \\
& -2g f^4_{abc} & :\partial_\mu u_a A^\mu_b \partial_\nu A^\nu_c : +
\nonumber \\
& -g f^5_{abc} & :\partial_\nu A^\nu_a u_b \partial_\mu A^\mu_c : \quad .
\label{var}
\end{eqnarray}
Gauge invariance requires that the gauge variation be a divergence:
\begin{eqnarray}
d_Q T_1 & = g\partial_\mu \bigl\{ & g^1_{abc} :\partial^\mu u_a
A_{\nu b} A^\nu_c : +
\nonumber \\
& & g^2_{abc} : u_a A^\mu_b \partial_\nu A^\nu_c : +
\nonumber \\
& & g^3_{abc} : \partial^\mu u_a u_b \tilde{u}_c : +
\nonumber \\
& & g^4_{abc} : u_a u_b \partial^\mu \tilde{u}_c : +
\nonumber \\
& & g^5_{abc} : \partial_\nu u_a A^\nu_b A^\mu_c : +
\nonumber \\
& & g^6_{abc} : u_a \partial^\mu A^\nu_b A_{\nu c} : +
\nonumber \\
& & g^7_{abc} : u_a \partial^\nu A^\mu_b A_{\nu c} : 
\bigr\} \quad , \label{div}
\end{eqnarray}
where $g^ 1_{abc}=g^ 1_{acb}, g^ 4_{abc}=-g^ 4_{bac}$.
Comparing the terms in (\ref{var}) and (\ref{div}), we immediately obtain
the following set of constraints for the coupling coefficients:
\begin{eqnarray}
-f^1_{cab} = 2 g^1_{abc} + g^6_{abc} & \mbox{from} & :\partial^\mu u
\partial_\mu A_\nu A^\nu :
\label{1} \\
g^2_{abc} + g^5_{abc} = -2f^4_{abc}
& & :\partial^\mu u A_\mu \partial_\nu A^\nu: \label{2} \\
g^2_{abc} + g^2_{acb} = -f^5_{bac}-f^5_{cab}
& & :u \partial_\nu A^\nu \partial_\mu A^\mu :
\label{3} \\
-f^2_{abc} = g^3_{abc} + 2 g^4_{abc} & & :\partial_\mu u u
\partial^\mu \tilde{u}:
\label{4} \\ 
g^3_{abc} = g^3_{bac} & & :\partial_\mu  u \partial^\mu u \tilde{u}: 
\label{5} \\
-f^1_{cba} - f^1_{bca} = g^5_{abc} + g^5_{acb} & & :\partial_\mu
\partial_\nu u A^\nu A^\mu:
\label{6} \\
-f^1_{acb} = g^5_{abc} + g^7_{abc} & &  :\partial_\mu u \partial_\nu
A^\mu A^\nu:
\label{7} \\
g^6_{abc}= - g^6_{acb} & & :u \partial_\mu A_\nu \partial^\mu A^\nu:
\label{8} \\
-f^2_{cab} = g^2_{acb} +g^7_{abc} & & :u \partial_\mu \partial_\nu
A^\mu A^\nu:
\label{9} \\
g^7_{abc}= - g^7_{acb} & &: u \partial_\nu A^\mu \partial^\mu A^\nu:
\label{10} \quad .
\end{eqnarray}
From (\ref{6}) (\ref{7}) and (\ref{10}) we immediately derive
\begin{equation}
f^1_{abc} + f^1_{acb} - f^1_{bca} - f^1_{cba} =0 \quad .
\end{equation}
Combining this with (\ref{asy}), 
we obtain the first important result that $f^1_{abc}$ is
totally antisymmetric.
$g^1_{abc}$ is symmetric in $b$ and $c$; from (\ref{8}) and (\ref{1}) we
conclude $g^1_{abc}=0$ and $g^6_{abc}=-f^1_{abc}$. 

The ansatz (\ref{div}) we used for $d_Q T_1$ is still too general,
because it has to fulfill $d_Q^2 T_1=0$. Calculating
$d_Q^2 T_1$ and setting the coefficients of all Wick monomials equal
to zero we get the relations
\begin{equation}
2g^ 1_{abc}-g^5_{bac}-g^ 5_{bca}+g^5_{abc}+g^6_{abc}+g^7_{abc}=0 \quad .
\end{equation}
$$2g^ 1_{abc}-g^ 5_{cba}+g^ 6_{abc}-g^ 7_{cba}=0$$
$$g^ 2_{abc}+g^ 2_{bac}+g^ 3_{abc}+g^ 3_{bac}+g^ 5_{abc}+g^ 5_{bac}=0$$
$$g^ 2_{abc}-g^ 3_{bac}-g^ 4_{bac}+g^ 4_{abc}+g^ 7_{acb}=0$$
$$g^ 6_{abc}+g^ 6_{acb}+g^ 7_{abc}+g^ 7_{acb}=0.$$
Combining these equations with the previous ones and taking all
symmetry properties of the $f$'s and $g$'s into account, we find
$f^ 2_{abc}=-f^ 1_{abc},\quad f^ 4=0,\quad f^ 5_{abc}=-f^ 5_{cba}$ 
(\ref{7})
\begin{equation}
g^ 1=0,\quad g^2_{abc} = -g^ 5_{abc}, \quad g^ 3=0, \quad 
g^4_{abc}=\frac{1}{2}f^1_{abc}, \quad g^ 6_{abc}=-f^ 1_{abc}
,\quad g^ 7_{abc}=f^ 1_{abc}-g^5_{abc}.
\end{equation}
Then 
$T_1$ assumes the following form
\begin{equation}
T_1=T_1^{YM} + T_1^{D} \quad ,
\end{equation}
\begin{equation}
T_1^{YM} = ig f^1_{abc} \bigl\{
:\frac{1}{2} A_{\mu a} A_{\nu b} F^{\nu \mu}_c:
-:A_{\mu a} u_b \partial^\mu \tilde{u}_c: \bigr\} \quad , \label{nrnr}
\end{equation}
\begin{equation}
T_1^{D} = igf^{5}_{abc}
:\partial_\nu A^\nu_a u_b \tilde{u}_c: \quad , \label{nr56}
\end{equation}
where 'YM' stand tentatively for 'Yang-Mills' and 'D' for 'Deformation', 
and $f^{5}_{abc}=-f^{5}_{cba}$.
\vskip 0.5 cm
\subsection{Second Order Gauge Invariance}
In this section we investigate the consequences of
second order gauge invariance for
the theory given by (\ref{nrnr})
\begin{equation}
T_1 = ig f_{abc} \bigl\{
:\frac{1}{2} A_{\mu a} A_{\nu b} F^{\nu \mu}_c:
-:A_{\mu a} u_b \partial^\mu \tilde{u}_c: \bigr\}
+ig\tilde{f}_{abc}:\partial_\nu A^\nu_a u_b \tilde{u}_c: \quad , \label{ym}
\end{equation}
where $f$ is totally antisymmetric and $\tilde{f}_{abc}$ antisymmetric in
$a$ and $c$.
If we consider the product
\begin{equation}
T_1(x)T_1(y) \quad ,
\end{equation}
then its gauge variation is simply given by
\begin{displaymath}
d_Q [T_1(x)T_1(y)] = d_Q T_1(x)T_1(y)
+T_1(x) d_Q T_1(y)
\end{displaymath}
\begin{equation}
= i \partial_\mu^x T_{1/1}^\mu(x)T_1(y)+iT_1(x)\partial_\mu^yT_{1/1}(y) \quad .
\end{equation}
But the analogous result
\begin{equation}
d_Q T_2(x,y) = i \partial_\mu^x T_{2/1}^\mu(x,y) + i \partial_\mu^y
T_{2/2}^\mu (x,y)
\end{equation}
is not automatically true. This can be seen as follows:
\begin{equation}
i \partial_\mu^x T_{1/1}^\mu (x)T_1(y) + (x \leftrightarrow y)
\end{equation}
contains terms
\begin{displaymath}
ig^2 \partial_\mu f_{abc} f_{a'b'c'}:u_a A_{\nu b} \partial^\mu A^\nu_c (x):
:A_{\mu' a'} A_{\nu' b'} \partial^{\nu'}A^{\mu'}_{c'}(y): + (x \leftrightarrow
y)
\end{displaymath}
\begin{eqnarray}
& = -g^2 \partial_\mu^x \bigl\{ & f_{abc}f_{cb'c'} :u_a(x)A_{\nu b}(x)
A_{\lambda b'} (y) \partial^\lambda A^\nu_{c'}(y): \partial^\mu_x D^+(x-y) +
\nonumber \\
& & f_{abc}f_{a'cc'} :u_a(x)A_{\nu b}(x)
A_{\lambda a'}(y) \partial^\nu A^\lambda_{c'}(y): \partial^\mu_x D^+(x-y) +
\nonumber \\
& & - f_{abc}f_{a'b'c} :u_a(x)A_{\nu b}(x)
A^\nu_{a'}(y) A^\lambda_{b'}(y): \partial^\mu_x \partial^\lambda_x D^+(x-y)
\bigr\}
\nonumber \\
& & + (x \leftrightarrow y) + ... \label{ord}
\end{eqnarray}
If we replace now the distribution $D^+(x-y)$ simply by the Feynman propagator
(time ordering of expression (\ref{ord})), we obtain additional local terms
('anomalies'), because $D_F$ satisfies the inhomogeneous wave equation
$\Box_x D_F(x-y) = \delta(x-y)$
\begin{eqnarray}
A &= -g^2 \bigl\{ & f_{abc}f_{cb'c'} :u_a(x)A_{\nu b}(x)
A_{\lambda b'}(y) \partial^\lambda A^\nu_{c'}(y): \delta(x-y) +
\nonumber \\
& & f_{abc}f_{a'cc'} :u_a(x)A_{\nu b}(x)
A_{\lambda a'}(y) \partial^\nu A^\lambda_{c'}(y): \delta(x-y) +
\nonumber \\
& & - f_{abc}f_{a'b'c} :u_a(x)A_{\nu b}(x)
A^\nu_{a'}(y) A^\lambda_{b'}(y):  \partial_\lambda^x \delta(x-y)
\bigr\}
\nonumber \\
& & + (x \leftrightarrow y) + ...
\end{eqnarray}
Using the distributional identity
\begin{displaymath}
A(x)B(y) \partial^x_\mu \delta(x-y) + A(y)B(x) \partial^y_\mu \delta(x-y) =
\end{displaymath}
\begin{equation}
A(x) \partial_\mu B(x) \delta(x-y) - \partial_\mu A(x) B(x) \delta(x-y) \quad ,
\end{equation}
we can rewrite A
\begin{eqnarray}
A &= -g^2 \bigl\{ & 2f_{abc}f_{cb'c'} :u_a(x)A_{\nu b}(x)
A_{\lambda b'}(y) \partial^\lambda A^\nu_{c'}(y): \delta(x-y) +
\nonumber \\
& &  2f_{abc}f_{a'cc'} :u_a(x)A_{\nu b}(x)
A_{\lambda a'}(y) \partial^\nu A^\lambda_{c'}(y): \delta(x-y) +
\nonumber \\
& &  f_{abc}f_{a'b'c} :u_a(x) \partial_\lambda A_{\nu b}(x)
A^\nu_{a'}(y) A^\lambda_{b'}(y): \delta(x-y)+
\nonumber \\
& &  f_{abc}f_{a'b'c} :\partial_\lambda u_a(x)  A_{\nu b}(x)
A^\nu_{a'}(y) A^\lambda_{b'}(y): \delta(x-y)+
\nonumber \\
& & - f_{abc}f_{a'b'c} :u_a(x)  A_{\nu b}(x)
\partial_\lambda A^\nu_{a'}(y) A^\lambda_{b'}(y): \delta(x-y)+
\nonumber \\
& & - f_{abc}f_{a'b'c} :u_a(x) A_{\nu b}(x)
A^\nu_{a'}(y) \partial_\lambda A^\lambda_{b'}(y): \delta(x-y) \Bigr\} + ...
\end{eqnarray}
Now we are free to add to $\partial_\lambda^x T_{2/1}^\lambda (x,y) +
(x \leftrightarrow y)$ a normalization term
\begin{displaymath}
-g^2 f_{abc} f_{a'b'c} \partial_\lambda^x \{ : u_a(x)A_{\nu b}(x) A^\nu_{a'}(y)
A^\lambda_{b'}(y): \delta(x-y)\} + (x \leftrightarrow y) =
\end{displaymath}
\begin{displaymath}
-g^2 f_{abc}f_{a'b'c} :\partial_\lambda u_a A_{\nu b} A^\nu_{a'} A^\lambda_{b'}
+u_a \partial_\lambda A_{\nu b} A^\nu_{a'} A^\lambda_{b'}+
\end{displaymath}
\begin{equation}
+u_a A_{\nu b} \partial_\lambda A^\nu_{a'} A^\lambda_{b'}
+\partial_\lambda u_a A_{\nu b} A^\nu_{a'} \partial_\lambda A^\lambda_{b'}:
\delta(x-y) \quad .
\end{equation}
Then the renormalized anomaly term $A'$ is given by
\begin{eqnarray}
A' &= -g^2  \bigl\{ & 2f_{abc}f_{cb'c'} :u_a(x)A_{\nu b}(x)
A_{\lambda b'}(y) \partial^\lambda A^\nu_{c'}(y): \delta(x-y) +
\nonumber \\
& &  2f_{abc}f_{a'cc'} :u_a(x)A_{\nu b}(x)
A_{\lambda a'}(y) \partial^\nu A^\lambda_{c'}(y): \delta(x-y) +
\nonumber \\
& &  2f_{abc}f_{a'b'c} :u_a(x) \partial_\lambda A_{\nu b}(x)
A^\nu_{a'}(y) A^\lambda_{b'}(y): \delta(x-y)+
\nonumber \\
& &  2f_{abc}f_{a'b'c} :\partial_\lambda u_a(x)  A_{\nu b}(x)
A^\nu_{a'}(y) A^\lambda_{b'}(y): \delta(x-y) \Bigr\} \quad . \label{can}
\end{eqnarray}
Furthermore, we must add a normalization term
\begin{equation}
N= i\frac{g^2}{2} f_{abc}f_{a'b'c'} : A_{\lambda a} A_{\nu b}
A^{\nu}_{a'} A^{\lambda}_{b'}: \delta(x-y) \label{viergl}
\end{equation}
to the gluon-gluon scattering term in $T_2(x,y)$ which contains
a propagator $\sim \partial^\mu \partial^\nu D_F(x-y)$.
Then the gauge variation of $N$
\begin{equation}
d_Q N = -2g^2 f_{abc}f_{a'b'c} :\partial_\lambda u_a
A_{\nu b} A^{\nu}_{a'} A^\lambda_{b'}: \delta(x-y)
\end{equation}
cancels the term in (\ref{can}) containing the operator $\partial_\lambda
u_a$.
The remaining anomalies cancel, iff
\begin{equation}
f_{abc}f_{dec}+f_{adc}f_{ebc}+f_{aec}f_{bdc}=0 \quad ,
\end{equation}
i.e. if the structure constants obey the {\it Jacobi identity}.
Consequently, the Lie algebra structure is obtained as a consequence of
perturbative gauge invariance at first and second order.

\subsection{Analysis of the Deformation}

Now we want to analyse the physical consequences of the various coupling
terms in (\ref{ym}) and (\ref{nr56}). 
The last term in (\ref{ym}) or (\ref{nrnr}) is a coboundary
because $\tilde f_{abc}$ is antisymmetric in $a$ and $c$:
\begin{equation}
\tilde f_{abc}:\d_\nu A^\nu_au_b\tilde u_c:=\frac{i}{2} d_Q
(\tilde f_{abc}:\tilde u_au_b\tilde u_c:)\> \quad . \label{nr72}
\end{equation}

Summing up we have shown that a gauge invariant coupling of gauge and
ghost fields is necessarily equal to the usual Yang-Mills form plus
a coboundary term. Now the question arises whether the latter
has physical consequences. Here the following result is relevant \cite{16}:
\vskip 0.2cm
{\bf Theorem} (Conjecture): Let $P$ be the projection operator on the physical
subspace (\ref{physspace})
and $T_n^0(x_1,\ldots,x_n)$ the $n$-point distribution for
the usual Yang-Mills theory (without divergence- and coboundary couplings).
If $T_n(x_1,\ldots,x_n)$ is the $n$-point distribution of the general
theory (with divergence- and coboundary couplings), then
\begin{equation}
PT_n(x_1,\ldots,x_n)P=PT_n^0(x_1,\ldots,x_n)P+{\mbox{(sum of
divergences)}} \quad . \label{nr78}
\end{equation}
\vskip 0.2cm
Unfortunately, the proof
is not complete in the case of coboundary couplings for arbitrary $n$
(it has been proven for $n\le 4$). But there is no doubt that the
theorem is true in general.

The sum of divergences in (\ref{nr78}) have no physical effect because they
would vanish in the formal adiabatic limit. Then (\ref{nr78}) means that the
physical S-matrix elements are unchanged.

\section{Massive Gauge Theories}

If some asymptotic fields are massive, their contribution to the gauge
charge (13) is changed according to
\begin{equation}
Q:=\int d^3x(\partial_\mu A^\mu_a (x)+m_a\Phi_a(x))  \stackrel
{\leftrightarrow}{\partial_0} u_a(x) \quad . \label{nr79}
\end{equation}
Here $\Phi_a(x)$ are scalar bosonic fields which are quantized according
to
\begin{equation}
(\w +m_a^2)\Phi_a(x)=0 \quad , \quad \>[\Phi_a(x),\Phi_b(y)]=-i\delta_{ab}
D_{m_a}(x-y) \quad . \label{nr80}
\end{equation}
This restores the nilpotency of $Q$.
The gauge variations (20) are altered as follows
\begin{equation}
[Q,\Phi_a]=im_au_a \quad , \quad \{Q,\tilde u_a\}=-i(\d_\mu A_a^\mu+
m_a\Phi_a) \quad . \label{nr81}
\end{equation}
The scalar fields $\Phi_a$ appearing in $Q$ do not belong to the
physical subspace (\ref{physspace}).

The starting point is now a general ansatz of the form
\begin{equation}
T_1=T_1^A+T_1^u+T_1^\Phi \quad , \label{nr82}
\end{equation}
where $T_1^A+T_1^u$ is the same as in (\ref{nr35}) and $T_1^\Phi$ is the most
general coupling of $\Phi_a$ and gauge and ghost fields. The discussion
of section 3.1 can be repeated, the only change are mass terms $\sim
m_a$ or $\sim m_a^ 2$ which come from the Klein-Gordon equation (\ref{nr80}) 
and from (\ref{nr81}).
The latter must be compensated by means of $T_1^\Phi$. This part
of the gauge invariance problem is considered in \cite{3} for the case of the
electroweak theory. The discussion of second order gauge invariance in
the Yang-Mills part (sect.3.2) remains unchanged because all anomalies
of sect.3.2 occur in the bigger theory as well. But there are many more
anomalies coming from $T_1^\Phi$. It turns out that it is impossible to
remove all those anomalies without enlarging the theory still further by
introducing another (now physical) scalar field. In this way the Higgs
field enters the scene. For the detailed discussion we refer to \cite{3,4}.


\vskip 1cm
\begin{thebibliography}{99}
\bibitem{1}
R. Stora, private communication
\bibitem{2}
A. Aste, M. D\"utsch, G. Scharf, J. Phys. {\bf {A}}: Math. Gen. 30, 5785 (1997).
\bibitem{3}
M. D\"utsch, G. Scharf, hep-th/9612091
\bibitem{4}
A. Aste, M. D\"utsch, G. Scharf, hep-th/9702053
\bibitem{5}
M. D\"utsch, T. Hurth, F. Krahe, G. Scharf, Nuovo Cimento {\bf {A106}},
1029 (1993), {\bf {A107}}, 375 (1994), {\bf {A108}}, 679 (1994).
\bibitem{6}
F. Krahe, Acta Physica Polonica {\bf {B27}}, 2453 (1996).
\bibitem{7}
N. N. Bogoliubov, D. V. Shirkov, {\em {Introduction to the theory of
quantized fields}} (Academic Press, New York, 1975).
\bibitem{8}
H. Epstein, V. Glaser, Ann. Inst. Poincar\'e {\bf {A19}}, 211 (1973).
\bibitem{9}
R. Stora, {\em {Differential algebras in lagrangean field theory}}
(ETH lectures, Z\"urich, 1994).
\bibitem{10}
G. Scharf, {\em {Finite quantum electrodynamics, the causal approach}}
(second edition, Springer, Berlin, Heidelberg, New York, 1995).
\bibitem{11}
T. Kugo, I. Ojima, Suppl. Prog. Theor. Phys. {\bf {66}}, 1 (1979).
\bibitem{12}
A. Aste, M. D\"utsch, G. Scharf, J. Phys. {\bf {A}}: Math. Gen. 31, 1563 (1998).
\bibitem{13}
J. C. Ward, Phys. Rev. {\bf {78}}, 182 (1950).
\bibitem{14}
A. V. Razumov, G. N. Rybkin, Nucl. Phys. {\bf {B332}}, 209 (1990).
\bibitem{15}
J. Bognar, {\em {Indefinite inner product spaces}} (Springer, Berlin, 1974).
\bibitem{16}
M. D\"utsch, J. Phys. {\bf {A}}: Math. Gen. 29, 7597 (1996).
\end{thebibliography}
\end{document}